\documentstyle[aps,twocolumn,psfig]{revtex}
\newcommand{\be}{\begin{equation}} 
\newcommand{\ee}{\end{equation}}
\newcommand{\bea}{\begin{eqnarray}}
\newcommand{\eea}{\end{eqnarray}}
\begin{document}
\draft
\title{ 
\begin{flushright}
\end{flushright}
\medskip
\bf {Hadronic Modes in the Quark Plasma with an Internal Symmetry }} 
\author
{ Munshi Golam Mustafa\thanks{Alexander von Humboldt Fellow, and on leave
from Saha Institute of Nuclear Physics, 1/AF Bidhan Nagar, Calcutta - 700 064,
India.}$^{,1,2}$, Abhijit Sen$^2$ and 
Lina Paria$^{3}$}

\medskip
\address
{$^1$ Institut f\"ur Theoretische Physik, Universit\" at Giessen, 
D-35392 Giessen, Germany \\
$^2$ Theory Group, Saha Institute of Nuclear Physics, 1/AF Bidhan Nagar, 
Calcutta-700 064, India\\
$^3$ Institute of Physics, Sachivalaya Marg, Bhubaneswar - 751 005, India}
\medskip
\medskip
\date{\today}
\maketitle

\begin{abstract}
\noindent We show that requiring the quark partition function
to be color singlet of $SU(3)$ color gauge group  leads to
reordering the thermodynamic potential in terms of the
colorless multi-quark modes ($q{\bar q}$, $qqq$, 
${\bar q}{\bar q}{\bar q}$, $\cdots$) at any given temperature.
These color singlet structures are not bound states in real sense,
rather they are combination of constituents quarks only. 
In accord with the $``$preconfinement" property of QCD, 
under a suitable confining mechanism, these could evolve into 
color singlet hadrons/baryons at low temperatures. At fairly 
high temperatures these multi-quark color singlet structures 
exist in the plasma like hadronic modes, just as in the 
more familiar low temperature phase. This suggests that there exists
a strong color correlation in the plasma at all temperatures.
\end{abstract}

\vskip 0.3in

{}
\narrowtext

\vskip 0.3in
The success of the quark model, the quantum chromodynamics (QCD), 
and the non-observability of the free partons ($q$, ${\bar q}$, 
$g$) has entailed the concept of confinement. QCD, the theory of 
strong interactions, is not perturbative at large distances. Thus, 
the confinement itself can not be treated perturbatively. 
There are reasons to believe that the confinement of partons
inside hadrons may not survive  collisions between heavy nuclei at 
relativistic energies \cite{harris}. One of the most interesting 
predictions of QCD at high temperature is the transition from the 
confined/chirally broken phase to deconfined/chirally symmetric state 
of quasi-free quarks and gluons, the so called quark-gluon plasma. 
At very high temperatures, the bulk properties (energy density, 
pressure, and entropy etc.) of QCD matter, seem to be described by a 
gas of nearly free quarks and gluons. However, it is also known that long 
range, non-perturbative effects disrupt this simple picture even at 
fairly high temperatures \cite{gross}.  

Lattice calculations 
\cite{detar} have provided ample evidence that the long distance
behaviour of high temperature phase is characterised by
the propagation of color singlet objects like multi-quark structures.
The determination of the plasma screening length shows evidence of 
a kind of correlations in the quark-gluon plasma at all temperatures. Now
what these structures are and how they show up is not yet very clear 
theoretically. 
Quite sometime ago an indication was made as a
$``$precursor" to the confinement property of QCD
by Amati and Veneziano \cite{amati} where the cascading and
fragmenting partons produced in hadronic collisions rearrange themselves
into color singlet clusters which ultimately evolve into hadrons
\cite{webber,klaus}. These considerations convince us that it is
important to incorporate the {\it dynamic} requirement of color 
singletness of the quark-matter to take into account such $`$interaction'
which $``$tunnels" into hadronic matter phase space \cite{klaus}.

The purpose of the present study is to reconsider  the statistical 
themodynamical description of quantum gases consisting of quarks and 
antiquarks, in such a way that the underlying symmetry can amount to 
reordering of thermodynamic potential in terms of the color singlet
multi-quark modes at any temperatures.
We note that the ingredients of our rather simple calculation have been
around for more than a decade\cite{red,elze,go,aub,mus}, but to our 
knowledge no-one has checked these dramatic consequences explicitly 
before. 

We begin with the partition function for a quantum gas, containing
quarks and antiquarks within a finite volume, which can be written as
\begin{equation}
{\cal Z} = {\rm{Tr}} \left ( {\hat {\cal P}} {\rm {exp}} (-\beta 
{\hat H}) \right ) \ \ , \label{part} 
\end{equation}
where $\beta=1/T$ is the inverse of temperature, ${\hat H}$ is the 
Hamiltonian of the physical system, and ${\hat {\cal P}}$ is the
projection operator to respect any configuration admitted by
a system. Now for a symmetry group ${\cal G}$ (compact Lie group) 
having unitary representation ${\hat U}(g)$ in a Hilbert space
${\cal H}$, the projection operator can be written as \cite{weyl}
\begin{equation}
{\hat{\cal P}}_j = d_j \int_{\cal G} {\rm d}\mu (g) \chi^{\star}_j(g)
{\hat U}(g) \ \ , \label{proj}
\end{equation}
where $d_j$ and $\chi_j$ are, respectively, the dimension and the
character of the irreducible representation $j$ of ${\cal G}$. 
${\rm d}\mu (g)$ is the normalised Haar measure in the group
${\cal G}$. The
symmetry group associated with the color singlet configuration of
the system is $SU(N_C)$, $N_C$ is the number of color corresponding to
fundamental representation. For the $SU(N_C)$ color singlet configuration
$d_j=1$ and $\chi_j=1$. The explicit form of the Haar measure corresponding
to $SU(N_C)$ can be found in Ref. \cite{aub} as
\begin{eqnarray}
\int_{SU(N_C)} {\rm d}\mu (g) &=& \frac{1}{N_C!} \left (\prod^{N_C-1}_{c=1} 
\int^\pi_{-\pi} \frac{{\rm d}\theta_c}{2\pi}\right ) \nonumber \\
&& \ \left [ \prod^{N_C}_{i < k} \left ( 2 \sin {\frac{\theta_i - \theta_k}
{2}}\right )^2 \right ] \ \ , \label{measure}
\end{eqnarray}
where $\theta_l$ is a class parameter obeys the preodicity condition 
$\sum^{N_C}_{l=1}\theta_l=0 \ ({\rm{mod}} 2\pi)$ which ensures that 
the group element is $SU(N_C)$. Now the partition function for the 
system becomes,
\begin{equation}
{\cal Z} = \int_{SU(N_C)} {\rm d}\mu (g) {\rm{Tr}} \left ( {\hat U}(g) 
{\rm {exp}} (-\beta {\hat H}) \right ) \ \ . \label{part1} 
\end{equation}

The Hilbert space ${\cal H}$ of the composite system has structure
of a tensor product of their individual Fock spaces as
\begin{equation}
{\cal H} = {\cal H}_q \otimes {\cal H}_{\bar q} \ \ , \label{prod}
\end{equation}
where the subscripts $q$ and $\bar q$ denote the quark and antiquark.
Now the trace involved in Eq.(\ref{part1}) decomposes into the product 
of two traces as
\begin{eqnarray}
{\cal Z} &=& \int_{SU(N_C)} {\rm d}\mu (g) \ {\rm{Tr}} \left ( {\hat U}_q(g) 
{\rm {exp}} (-\beta {\hat H}_q) \right ) \nonumber \\  
&& \ \ \ \ \ \ \ \ \ \ \ \ {\rm{Tr}} \left ( {\hat U}_{\bar q}(g) 
{\rm {exp}} (-\beta {\hat H}_{\bar q}) \right ) \ \ . \label{decom} 
\end{eqnarray}
The traces involved in Eq.(\ref{decom}) can now be evaluated using the 
known results for fermions (for details see Ref.\cite{aub})and it becomes
\begin{equation}
{\cal Z} = \int_{SU(N_C)} {\rm d}\mu (g) \exp\left (\Theta\right ) \ \ ,
\label{part2}
\end{equation}
with
\begin{eqnarray}
\Theta &=& {\rm{tr}} \big [ \ln\left (1+e^{i\theta_c} e^{-\beta
(\epsilon^q_{\alpha}-\mu^q)} \right ) \ \nonumber \\
&+& \ln\left (1+e^{-i\theta_c} e^{-\beta (\epsilon^{\bar q}_{\alpha}-
\mu^{\bar q})} \right ) 
\big ] \ \ \ , \label{theta}
\end{eqnarray}
where trace indicates the summation over color (c), flavour (q), spin (s) 
and single particle states ($\alpha$) with $\epsilon_{\alpha}= \sqrt{
p^2_{\alpha}+m^2}$. $\mu^q$ and $\mu^{\bar q}$ are, respectively, quark
and antiquark chemical potential. In Eq.(\ref{theta}) one can replace
$\epsilon^{\bar q}$ by $\epsilon^q$, and $\mu^{\bar q}$
by $-\mu^q$. It is worth noting here that Eq.(\ref{part2}) represents
the general structure of color projected partition function so far existing 
in the literature \cite{red,elze,go,aub,mus}, and 
depending upon the nature of the problem it has been utilised 
accordingly.  Now for convenience, we make a substitution $\xi^q = -i\beta
\mu^q$ and after a little algebra, Eq.(\ref{theta}) becomes
\begin{eqnarray}
\Theta &=& \ln\Big \{ \prod_{\alpha} \prod^{N_C}_c \prod^{N_f}_q \prod^{N_s}_s 
\Big [e^{-\beta \epsilon^q_{\alpha}}  \ \times \nonumber \\
&&  \ \ \ \ \left ( 2 \cosh \beta \epsilon^q_{\alpha}
+2 \cos (\theta_c +\xi^q) \right ) \Big ] \Big \} . \label{simpl}
\end{eqnarray}
where $N_f$ and $N_s$, are, respectively the number of flavor and 
spin degrees of freedom. On substitution Eq.(\ref{simpl}) in 
Eq.(\ref{part2}), it becomes for a finite volume
\begin{eqnarray}
{\cal Z} &=& \prod_{\alpha} \int_{SU(N_C)} {\rm d}\mu (g) 
\Big \{  \prod^{N_C}_c \prod^{N_f}_q \prod^{N_s}_s 
\Big [e^{-\beta \epsilon^q_{\alpha}}  \ \times \nonumber \\
&&  \ \ \ \ \left ( 2 \cosh \beta \epsilon^q_{\alpha}
+2 \cos (\theta_c +\xi^q) \right ) \Big ] \Big \} . \label{part3}
\end{eqnarray}
We would like to note that the product over $\alpha$ is written outside 
the group integration without any loss of generality. For exact flavor
symmetry Eq.(\ref{part3}) can be written as
\begin{eqnarray}
{\cal Z} &=& \prod_{\alpha} \int_{SU(N_C)} {\rm d}\mu (g) 
\Big \{  \prod^{N_C}_c  
\Big [e^{-\beta \epsilon_{\alpha}}  \ \times \nonumber \\
&&  \ \ \ \ \left ( 2 \cosh \beta \epsilon_{\alpha}
+2 \cos (\theta_c +\xi) \right ) \Big ]^{2N_f} \Big \} . \label{part4}
\end{eqnarray}
For $N_C=3$ and $N_f=2$, the above equation becomes,
\begin{eqnarray}
{\cal Z} &=& \prod_{\alpha} \int_{SU(3)} {\rm d}\mu (g) 
\Big \{  2^{12}e^{-12\beta\epsilon_\alpha} \nonumber \\ 
&& \ \ \ \ \times \left [ \cosh  \beta \epsilon_\alpha + 
\cos \left (\theta_1 +\xi \right ) \right ]^4 \nonumber \\
&& \ \ \ \ \times \left [ \cosh  \beta \epsilon_\alpha + \cos 
\left (\theta_2 +\xi \right ) \right ]^4 \nonumber \\
&& \ \ \ \ \times \left [ \cosh  \beta \epsilon_\alpha + \cos 
\left (\theta_1+\theta_2 -\xi \right ) \right ]^4 \Big \} \ \ , \label{part5}
\end{eqnarray}
with the measure corresponding to $SU(3)$ color symmetry can be 
obtained from Eq.(\ref{measure})
as
\begin{eqnarray}
\int_{SU(3)}{\rm d}\mu(g) &=& \frac{8}{3\pi^2} \int^{\pi}_{-\pi} 
{\rm d}\theta_1
\int^{\pi}_{-\pi} {\rm d}\theta_2 \ \sin^2 {\frac {\theta_1 -\theta_2} {2}}
\nonumber \\
&& \times \sin^2 {\frac {\theta_1 +2\theta_2} {2}} \ 
\sin^2 {\frac {2\theta_1 +\theta_2} {2}} \ \ , \label{measure1}
\end{eqnarray}
where we have made use of periodicity condition 
$\sum^{N_C=3}_{l=1} \theta_l=0$. Substituting Eq.(\ref{measure1}) in 
Eq.(\ref{part5}), and performing the several hundred elementary integrations
in the group space, one can write the color singlet thermodynamic 
potential for finite system as
\begin{equation}
\Omega = -\frac {1}{\beta} \ \sum_{\alpha} \ \ln \Big [ 1 + M + B
\Big ] \  \ , \label{therpot}
\end{equation}
where $M$ corresponds to mesonic modes and is given as
\begin{eqnarray}
M &=& 16 e^{-2\beta\epsilon_\alpha} 
+ 136 e^{-4\beta\epsilon_\alpha} 
+ 816 e^{-6\beta\epsilon_\alpha} 
+ 1616 e^{-8\beta\epsilon_\alpha} 
\nonumber \\
&+& 4941 e^{-10\beta\epsilon_\alpha} 
+ 6160 e^{-12\beta\epsilon_\alpha} 
+ 4941 e^{-14\beta\epsilon_\alpha} 
\nonumber \\
&+& 1616 e^{-16\beta\epsilon_\alpha} 
+ 816 e^{-18\beta\epsilon_\alpha} 
+ 136 e^{-20\beta\epsilon_\alpha} 
\nonumber \\
&+& 16 e^{-22\beta\epsilon_\alpha} 
+ e^{-24\beta\epsilon_\alpha} \ \ \ , \label{meson}
\end{eqnarray}
whereas those of baryonic (antibaryonic) modes are obtained as
\begin{eqnarray}
B &=& 20 e^{-3\beta (\epsilon_{\alpha}\mp\mu)} 
+180 e^{-2\beta \epsilon_\alpha} \ e^{-3\beta (\epsilon_{\alpha}\mp\mu)} 
\nonumber \\
&+& 816 e^{-4\beta\epsilon_\alpha} \ e^{-3\beta (\epsilon_{\alpha}\mp\mu)} 
+ 2320 e^{-6\beta\epsilon_\alpha} \ e^{-3\beta (\epsilon_{\alpha}\mp\mu)} 
\nonumber \\
&+& 3020 e^{-8\beta\epsilon_\alpha} \ e^{-3\beta (\epsilon_{\alpha}\mp\mu)} 
+3020 e^{-10\beta\epsilon_\alpha} \ e^{-3\beta (\epsilon_{\alpha}\mp\mu)} 
\nonumber \\
&+& 2320 e^{-12\beta\epsilon_\alpha} \ e^{-3\beta (\epsilon_{\alpha}\mp\mu)} 
+816 e^{-14\beta\epsilon_\alpha} \ e^{-3\beta (\epsilon_{\alpha}\mp\mu)} 
\nonumber \\
&+& 180 e^{-16\beta\epsilon_\alpha} \ e^{-3\beta (\epsilon_{\alpha}\mp\mu)} 
+20 e^{-18\beta\epsilon_\alpha} \ e^{-3\beta (\epsilon_{\alpha}\mp\mu)} 
\nonumber \\
 &+& 50 e^{-6\beta (\epsilon_{\alpha}\mp\mu)} 
+240 e^{-2\beta \epsilon_\alpha} \ e^{-6\beta (\epsilon_{\alpha}\mp\mu)} 
\nonumber \\
&+& 570 e^{-4\beta\epsilon_\alpha} \ e^{-6\beta (\epsilon_{\alpha}\mp\mu)} 
+ 800 e^{-6\beta\epsilon_\alpha} \ e^{-6\beta (\epsilon_{\alpha}\mp\mu)} 
\nonumber \\
&+& 570 e^{-8\beta\epsilon_\alpha} \ e^{-6\beta (\epsilon_{\alpha}\mp\mu)} 
+ 240 e^{-10\beta\epsilon_\alpha} \ e^{-6\beta (\epsilon_{\alpha}\mp\mu)} 
\nonumber \\
 &+& 50 e^{-12\beta\epsilon_\alpha} \ e^{-6\beta (\epsilon_{\alpha}\mp\mu)} 
\nonumber \\
&+& 20 e^{-9\beta (\epsilon_{\alpha}\mp\mu)} 
+ 40 e^{-2\beta\epsilon_\alpha} \ e^{-9\beta (\epsilon_{\alpha}\mp\mu)} 
\nonumber \\
&+& 40 e^{-4\beta\epsilon_\alpha} \ e^{-9\beta (\epsilon_{\alpha}\mp\mu)} 
+ 20 e^{-6\beta\epsilon_\alpha} \ e^{-9\beta (\epsilon_{\alpha}\mp\mu)} 
\nonumber \\
&+& e^{-12\beta (\epsilon_{\alpha}\mp\mu)}  \ \ . \label{baryon}
\end{eqnarray}
In the infinite volume limit the $\sum_\alpha$ in Eq.(\ref{therpot}) 
can be replaced by integration as
\begin{equation}
\Omega = -\frac{1}{\beta} \ \int \frac {{\rm d}^3p}{(2\pi)^3} \  
\ln \left [ 1+M+B \right ] \ \ . \label{therpot1}
\end{equation}
It is to be noted here that in Eqs.(\ref{meson}) and (\ref{baryon}) one 
needs to replace $\epsilon_\alpha$ by just $\epsilon$.

 Equation(\ref{therpot1}) clearly shows that the color projection amounts 
to reordering the thermodynamic potential in terms of Boltzmann factors 
for the colorless multi-quark (mesonic/baryonic) modes at any temperature.
The mesonic modes can be seen from Eq.(\ref{meson}) having quark content
$q^{n} {\bar q}^{n}, \ n=1, \cdots, 12$ with energy $2n\epsilon$. On
the other hand baryonic (antibaryonic) modes are very transparent from 
Eq.(\ref{baryon}) with quark content $q^{m+N_CB}{\bar q}^m \ \ 
({\bar q}^{m+N_C{\bar B}} q^m)$; $B ({\bar B})$=1, 2, 3, 4, is baryon 
(antibaryon) number, and $m=0,1, \cdots$ (restricted by the values of
$B ({\bar B})$, e.g., see Eq.(\ref{baryon})), and $N_C=3$.  In 
general the energy of these baryonic (antibaryonic) modes can be written as 
$2m\epsilon +B N_C (\epsilon \mp \mu)$. The maxium values of $n$, $m$ and
$B({\bar B})$ depend on the number of flavor chosen ($e.g.$, see 
Eq.(\ref{part4})).  The interesting feature of this color
projection is that the chemical potential, $\mu$, always appears with the color 
factor $N_C$ in the baryonic Boltzmann factor. Of-course, these color singlet
structures are not bound states in real sense, rather we should say that
they are combination of constituents quarks only. 

Under a suitable confining mechanism, we hope that these multi-quark 
colorless structures could evolve into color singlet hadrons in the low
temperature limit. This is in accord with the $``$preconfinement" property
of QCD noted by Amati and Veneziano \cite{amati} quite sometime ago.
In the mesonic sector, $n=1$ corresponds to low lying mesons (first term
in Eq.(\ref{meson})) whereas $n>1$ represents exotic mesons. It is to be
noted that the factor of 16 appearing in the first term in Eq.(\ref{meson}) 
can amount to 16 low lying degenerate mesonic states corresponding to $SU(2)$  
flavor and $SU(2)$ spin symmetry. We would like to point out here that
the color singlet thermodynamic potential can possibly extract the mesonic
states reasonably well in  which pions, being the pseudoscalar Goldstone boson, 
could be the exception. On the other hand $m=0$, $B({\bar B})=1$ and
$N_C=3$ amount to  low lying baryons (first term in Eq.(\ref{baryon}))
whereas $m>1$, $B({\bar B}) \ge 1$ and $N_C=3$ correspond to exotic 
baryons. The factor of 20 appearing in the first term in Eq.(\ref{baryon}) 
representing 20 baryonic and 20 antibaryonic states. 
This is also quite consistent with $SU(2)$ flavor and $SU(2)$ spin symmetry 
in which nucleons and deltas are degenerate, and their total degeneracy is 40. 
It is also very transparent that, as expected physically, the low 
lying hadronic modes play a dominant role at low temperature, while the exotic 
hadronic modes/collective modes are most relevant at high temperature.  
In a nonsupersymmetric tachyonless string model, Kutasov and Seiberg
\cite{kuta} suggested that the number of fermionic and bosonic states must 
approach each other as increasingly massive states are included in the 
hadronic
density of states. On the basis of this result, Freund and Rosner 
\cite{freund} proposed that there must exist exotic mesons and baryons
with similar quark content as discussed above to have equalization 
of mesonic and baryonic density of states since the observed states 
are deficient at a higher mass range (for meson it is 1.3 GeV and for
baryon it is 2 GeV).  In the low temperature limit the existence
of these exotic hadrons is highly improbable, but they might start
appearing with the increase of temperatures. In a model dependent 
calculation \cite{mus1} the limiting temperature has been found to 
vary with the mass of the hadrons.

As discussed above, those multi-quark color singlet modes become
very relevant at very high temperatures. The estimation of energy
density and pressure \cite{go,mus} at moderately high temperatures
show deviation from their ideal gas behaviour. In this context, it
has been suggested \cite{detar} that the long distance behaviour of 
the high temperature phase can be characterised by the propagation 
of color
singlet multi-quark structures. In this simple minded calculation
such color singlet objects appear very cleanly at high temperature.
Now the question comes what these modes correspond
at high temperature. 
As we see these multi-quark objects are not real bound states,
but they are just combination of constituent quarks
and indistinguishable from the color singlet
hadrons at low temperature. This could lead to a kind of analog of the 
high temperature phase to that of low temperature, and  we naively 
speculate that if there is a phase transition in light-quark QCD, 
it may be of chiral character than a deconfinement character. However,
this issue requires a more careful investigation.

Finally, we would like to make a comment here on one important aspect 
of three flavor case. If one considers three flavors (calculation will 
really be very combursome), there
will be a structure consisting of 6-quarks (two of each flavor) which 
could be of particular interest to the community who are searching
for strangelets in heavy ion collisions. This particular structure 
is known as quark-alpha ($Q_{\alpha}$) in the literature\cite{michel} 
which could be a probable candidate for detecting stranglets in heavy 
ion collisions.

\vskip 0.3 in 

The authors are thankful to M. Thoma, K. Redlich and D. K. Srivastava for
helpful discussion during the course of this work. MGM gratefully  
acknowledges the financial support from the Alexander von Humboldt 
Foundation, Bonn, Germany.

\end{document}